\documentclass[12pt,letterpaper]{article}

\usepackage{epsfig}
\usepackage{amsmath}
\usepackage{amssymb}
\usepackage{amsthm}
\usepackage{indentfirst}
\usepackage{xspace}
\usepackage{multirow}
\usepackage{hyperref}
\usepackage{xcolor}
\definecolor{darkred}{rgb}{0.5,0.15,0.15}
\hypersetup{colorlinks=true,urlcolor=darkred,linkcolor=darkred,citecolor=darkred}
\usepackage{verbatim}
\usepackage[letterpaper,margin=1in,headheight=15pt]{geometry}
\usepackage{mathpazo}
\usepackage{authblk}
\usepackage{tikz-cd}
\usepackage{capt-of}
\usepackage{silence}
\usepackage[missing=]{gitinfo2}

\usepackage{tikz}
\usetikzlibrary{calc}   
\usetikzlibrary{topaths}
\usetikzlibrary{decorations}
\usetikzlibrary{decorations.pathmorphing}

\numberwithin{equation}{section}

\newcommand{\hC}{{\widehat{C}}}

\newcommand{\fg}{{\mathfrak g}}

\newcommand{\cW}{\ensuremath{\mathcal W}}

\newcommand{\R}{\ensuremath{\mathbb R}}
\newcommand{\C}{\ensuremath{\mathbb C}}
\newcommand{\PP}{\ensuremath{\mathbb P}}
\newcommand{\Z}{\ensuremath{\mathbb Z}}

\newcommand{\bbS}{\ensuremath{\mathbb S}}

\newcommand{\half}{\ensuremath{\frac{1}{2}}}

\newcommand{\N}{{\mathcal N}}

\newcommand{\WW}{{\mathcal{W}}}

\newcommand{\ti}[1]{\textit{#1}}

\DeclareMathOperator{\Tr}{Tr}

\begin{document}

\title{Factorized class $S$ theories and surface defects}
\date{{{\tiny \color{gray} \tt \gitAuthorIsoDate}}
{{\tiny \color{gray} \tt \gitAbbrevHash}}}
\author[1]{Behzat Ergun}
\author[2]{Qianyu Hao}
\author[3]{Andrew Neitzke}
\author[4]{Fei Yan}
\affil[1,2]{Department of Physics, University of Texas at Austin}
\affil[3]{Department of Mathematics, Yale University}
\affil[4]{NHETC and Department of Physics and Astronomy, Rutgers University}

\maketitle

{\abstract{It is known that some theories of class $S$ are actually factorized
into multiple decoupled nontrivial four-dimensional $\N=2$ theories. We propose a way of
constructing examples of this phenomenon using the physics of half-BPS 
surface defects, and check that it works in one simple example:
it correctly reproduces a known realization of two copies of $\N=2$ superconformal $SU(2)$ QCD,
describing this factorized theory 
as a class $S$ theory of type $A_3$ on a five-punctured sphere with a twist line.
Separately, we also present explicit checks
that the Coulomb branch of a putative factorized class $S$ theory 
has the expected product structure, in two examples.}}

\setcounter{page}{1}


\section{Introduction}

\subsection{Factorized class \texorpdfstring{$S$}{S} theories}

The class $S$ theory $S[\fg,C]$ is the twisted compactification of a six-dimensional
$(2,0)$ theory of type $\fg$ 
on a Riemann surface $C$, with appropriate decorations (e.g. codimension-2 defects or twist lines)
inserted on $C$, in the
limit where the volume of $C$ is taken to zero. It is believed that
one obtains in this way 
a four-dimensional $\N=2$ supersymmetric field theory depending on the conformal structure
of $C$.
It has turned out to be possible to realize a large number of
four-dimensional $\N=2$ theories in this way, including conventional gauge theories
as well as apparently non-Lagrangian theories; in either case, the six-dimensional
perspective has turned out to give useful insights into the dynamics of the theory.

\begin{figure}
	\centering
	\includegraphics[width=100pt]{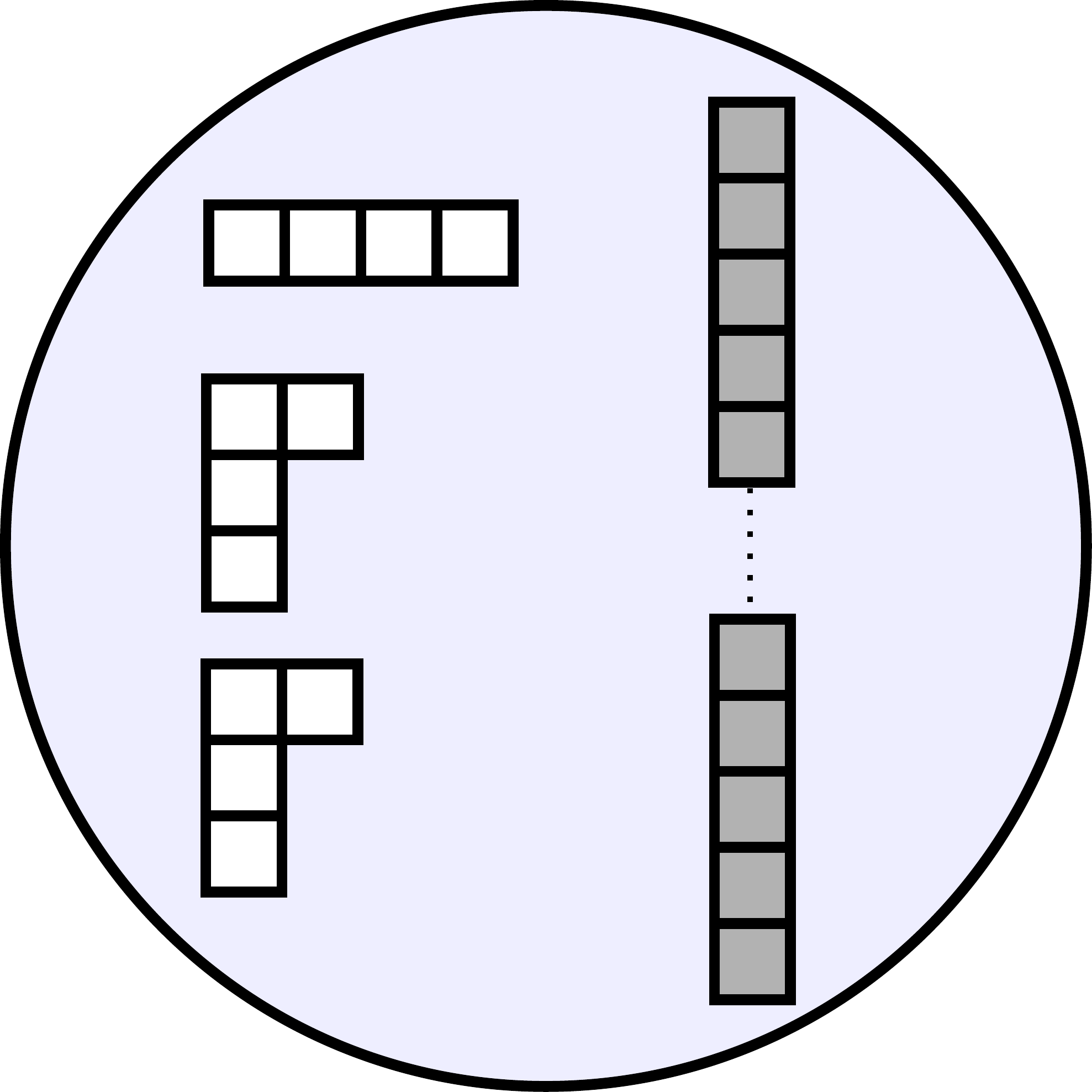}
	\caption{A five-punctured sphere $C$ for which the theory
	$S[A_3, C]$ is factorized. $C$ contains one maximal puncture (represented by the Young diagram $[1^4]$), two minimal punctures (represented by the Young diagram $[3,1]$), and two minimal twisted punctures (represented by the Young diagram $[5]$). There is a $\mathbb{Z}_2$ twist line (black dotted line) connecting the 
	two twisted punctures.}
	\label{fig:example-C}
\end{figure}

It was pointed out in \cite{Chacaltana:2012ch} 
that, in the landscape of $\N=2$ theories of class $S$, 
there are some theories which factorize into 
direct sums of decoupled constituents.
For example, suppose we take $\fg = A_3$, and take
$C$ to be a sphere with one maximal puncture, two minimal punctures and two minimal twisted punctures (here ``twisted'' refers to a twist by the $\Z_2$
outer automorphism of $A_3$), as shown in \autoref{fig:example-C}.
According to \cite{Chacaltana:2012ch}, the theory $S[\fg,C]$ 
is factorized: it is
the sum of two decoupled copies of the 
$SU(2)$ gauge theory with $N_f=4$ fundamental hypermultiplets.

This factorization might seem surprising from the six-dimensional 
point of view; there is
no obvious reason why this particular combination of punctures ought to 
lead to a factorized theory. Many other examples of the same general
phenomenon were discovered in  \cite{Chacaltana:2011ze,Chacaltana:2014jba,Chacaltana:2015bna,Distler:2017xba,Distler:2018gbc,Chacaltana:2017boe,Chcaltana:2018zag}.

\subsection{Why do factorized class \texorpdfstring{$S$}{S} theories exist?}

The purpose of this paper is to suggest a partial explanation for 
why these factorized class $S$ theories exist.
The explanation relies on an alternative point of view on the class $S$
construction, as follows. 

In the six-dimensional $(2,0)$ theory 
there exists a distinguished $\half$-BPS surface defect.\footnote{or more precisely one
such defect for every representation of the Lie algebra labeling the theory; in 
type $A$ we can just take the fundamental representation.} 
Placing this defect at a point $z \in C$ yields a $\half$-BPS surface defect $\bbS$
in the theory $S[\fg,C]$, sometimes called
the \ti{canonical surface defect}.
The space of marginal chiral deformations of the defect $\bbS$ is $C$.
Thus the surface $C$ has a dual meaning: 
on the one hand it is the surface on which one compactifies to construct the theory $S[\fg,C]$;
on the other hand is the moduli space parameterizing marginal
chiral deformations of the defect $\bbS$
in the theory.

This leads to an approach to constructing a class $S$ realization of one's
favorite $\N=2$ field theory: one should look for a $\half$-BPS 
surface defect in the theory, whose space of marginal chiral deformations
is a complex curve $C$.
If the surface defect has $k$ vacua, then the Lie algebra $\fg$ should admit
a $k$-dimensional representation; the simplest choice (and the one which works
in the example we study in this paper) would be $\fg = A_{k-1}$.
By studying the chiral ring of the surface defect
as a function of the moduli, as we will discuss below, 
one can detect what sorts of punctures $C$
should carry. 
Then, the proposal is that the $\N=2$ theory we started with might be 
realized as the class $S$ theory $S[\fg,C]$.

In particular, we can apply this proposal in the case where
the $\N=2$ theory we start with is factorized. 
Then we can consider a surface defect which couples to all
of the factors (thus indirectly coupling the different factors along
the defect worldvolume), and in this way produce a candidate for a class $S$ realization
of the factorized theory.

\subsection{An example}
\label{intro:an example}

As a test of this strategy, we looked to see whether we could use it 
to recover the simplest known example, the class $S$ construction of the doubled 
$SU(2)$ $N_f=4$ theory which we recalled above.
For this purpose the first step is 
to exhibit a $\half$-BPS surface defect $\bbS$
in this theory, whose moduli space is the curve $C$
of \autoref{fig:example-C}.

One way to build $\half$-BPS surface defects in an $\N=2$ gauge theory
with gauge group $G$ is to take a
two-dimensional $\N=(2,2)$ theory on which $G$ acts by flavor symmetries.
For instance one could take
a supersymmetric sigma model on a K\"ahler target $X$, with $G$ acting
by isometries on $X$. 
In our case we have
\begin{equation}
	G = SU(2) \times SU(2)
\end{equation}
and since we want the surface defect to have a 1-dimensional
moduli space we should arrange that $h^{1,1}(X) = 1$. 
One natural candidate, then, would be to take 
\begin{equation}
	X = \C\PP^3
\end{equation}
with $G$ as the subgroup $SU(2) \times SU(2) \subset SU(4)$.
It turns out that this surface defect is not exactly the one 
we are looking for; it cannot be, since it does not break the bulk
$SO(8) \times SO(8)$ flavor symmetry, while the canonical surface
defect in the class $S$ realization we are after does break this
flavor symmetry.
Thus we consider a slight variant described in \cite{Gaiotto:2013sma}, 
a gauged linear sigma model augmented by extra couplings to the bulk fields 
which give the desired flavor symmetry breaking.

Once we have identified our candidate surface defect we need to understand
its chiral ring, as a function of the coupling of the defect. 
Fortunately, technology for studying surface
defect chiral rings in Lagrangian $\N=2$ theories has already been
developed in \cite{Gaiotto:2013sma}, and their results can be applied
directly in our situation. When we do so, we find a parameter space
$\hC$ which is not quite the one we were hoping for:
in particular, on $\hC$ we have $6$ punctures instead of $5$. 
However, $\hC$ has a natural 
$\Z_2$ symmetry, which we interpret as a duality symmetry of the 
surface defect theory. With this interpretation
the parameter space of inequivalent surface defects is the 
quotient
\begin{equation}
 C = \hC / \Z_2
\end{equation}
The $\Z_2$ action identifies the $6$ punctures in pairs,
giving $3$ ordinary (untwisted) punctures on $C$.
The $2$ fixed points of the $\Z_2$ become ramification points
for the covering $\hC \to C$; their images on $C$ 
are interpreted as twisted punctures.
In this way we obtain the expected list of $5$ punctures on $C$.
Moreover, we find that the relation between the two 
complex structure moduli 
of $C$ and the two $SU(2) \times SU(2)$ 
gauge couplings is just as expected from \cite{Chacaltana:2012ch}.

Thus, at least in this example, our strategy for finding 
class $S$ realizations of factorized theories works.

\subsection{Comments and future directions}

\begin{itemize}

\item
Our computation gives a ``proof of concept'' but not a full explanation 
of the existence of factorized class
$S$ theories, since we have not proven that a theory admitting a surface 
defect with 1-dimensional moduli space $C$ is necessarily 
of the form $S[\fg,C]$.
This leads to the question: can we identify necessary and sufficient 
conditions under which the theory really is $S[\fg,C]$?

One condition which is surely necessary is that the surface defect
is coupled nontrivially to all of the decoupled factors of the theory.
A sharper version of this condition is as follows.
Given a surface defect with moduli space $C$,
there is a map from the
conformal manifold of the 4d theory
to the space of complex structures on $C$.
If the theory is actually $S[\fg,C]$, then this map should be
a local isomorphism; checking this local condition 
boils down to studying the OPE between bulk
chiral operators and boundary anti-chiral operators \cite{nstoappear}, which could 
be analyzed in specific examples.
(Note that there are examples of theories $S[\fg,C]$
for which it is believed that the conformal manifold of $S[\fg,C]$ is 
not equal to the space of 
complex structures on $C$, but rather is a finite covering of it; indeed
the main example we consider in this paper is of that sort, as noted in
\cite{Chacaltana:2012ch}. Thus we should not expect to replace
``local isomorphism'' by ``global isomorphism'' above.)

\item
Although we only studied the simplest example of a factorized class $S$ theory 
here, there are various other examples which should also 
fit into our framework. For instance,  
\cite{Chacaltana:2012ch} gives a class $S$ construction 
of the doubled $SU(N)$ $N_f = 2N$ theory, 
which should be realizable using a slight generalization of
the surface defect construction we give here. We expect that
many more candidates for factorized theories, including theories with arbitrarily
many constituents, could be constructed using the
same approach.

\item
More generally, there are also examples of factorized class $S$ theories
where the constituent theories are non-Lagrangian. For example, \cite{Chacaltana:2011ze} 
gives a class $S$ theory which turns out to be factorized into two 
copies of the $E_6$ Minahan-Nemeschansky theory (see \autoref{fig:D4eg} below). 
Thus there
should be a surface defect which couples those two theories and which
has a 1-dimensional moduli space. It would be interesting to come up with
a way of constructing that surface defect directly, and thus account for
the existence of this factorized class $S$ theory.

\end{itemize}

\subsection{Factorization on Coulomb branches}\label{sec:CBfactorize}

When an ${\N=2}$ theory factorizes into a direct sum of simpler constituents, 
its moduli space must factorize as the product of the moduli spaces of the constituent theories. 
For the class $S$ theories we are discussing, direct evidence for 
factorization of the Higgs branches has been seen in \cite{Chacaltana:2015bna,Yan2018}, by computing appropriate limits of the superconformal index \cite{Gadde_2012}. However, factorization of the 
Coulomb branches has not been investigated directly.
In \autoref{chapter:CBfactorization} below, we briefly discuss the Coulomb branch factorization
as follows.

In an $\N=2$ theory factorized into two constituents,
the couplings must split as $q = (q_1,q_2)$, 
the Coulomb branch moduli must split
as $u = (u_1, u_2)$, and the lattice of charges must split as
$\Gamma = \Gamma_1 + \Gamma_2$.
Moreover, the periods $Z_\gamma(q,u)$ 
of the Seiberg-Witten differential must depend only on the appropriate moduli:
if we write a charge with respect
to this decomposition as $\gamma = (\gamma_1, \gamma_2)$, we must have
\begin{equation}\label{eqn:Zgammadecom}
	Z_\gamma(q,u) = Z_{\gamma_1}(q_1,u_1) + Z_{\gamma_2}(q_2,u_2).
\end{equation}
Even when this decomposition exists, it is not generally obvious at first look.
We check this decomposition (in specific regions of the Coulomb branch)
in two examples: the factorized theory associated to \autoref{fig:example-C} (doubled
$SU(2)$ $N_f=4$ theory)
and the one in \autoref{fig:D4eg} below (doubled $E_6$ Minahan-Nemeschansky theory).

\subsection*{Acknowledgements}

We would like to thank Jacques Distler and Pietro Longhi for very helpful discussions. We thank Jacques Distler for very helpful comments on a draft.
BE is partially supported by NSF grant PHY-1914679.
AN's work on this project was supported by NSF grants DMS-1711692
and DMS-2005312.
FY is supported by DOE grant DE-SC0010008.

\section{Warm-up: \texorpdfstring{$SU(2)$}{SU(2)} with \texorpdfstring{$N_f=4$}{Nf=4}}\label{onecopy}

In this section we briefly recall several ways of thinking about surface defects,
and their chiral rings, on the Coulomb branch of the $\N=2$ $SU(2)$ theory with $N_f=4$.

\begin{figure}
\centering
\includegraphics[width=250pt]{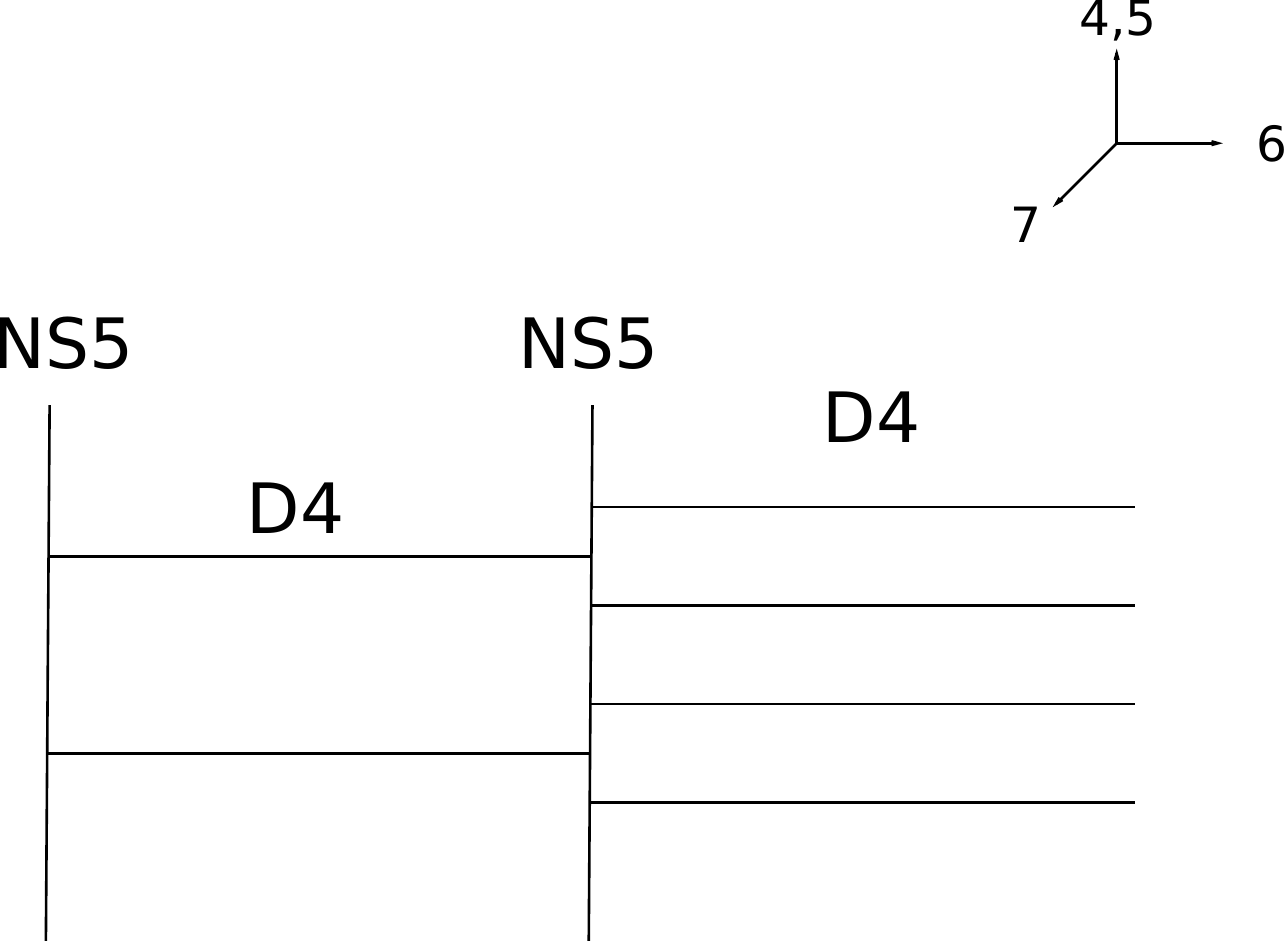}
\caption{A brane construction for  \texorpdfstring{$SU(2)$}{SU(2)} with \texorpdfstring{$N_f=4$}{Nf=4}, where all the four semi-infinite D4-branes end on one NS5-brane. This brane construction corresponds to the asymmetric setup discussed in \autoref{sec:asymbraneSU2single}.}
\label{fig:asym1copy}
\end{figure}

\begin{figure}
\centering
\includegraphics[width=250pt]{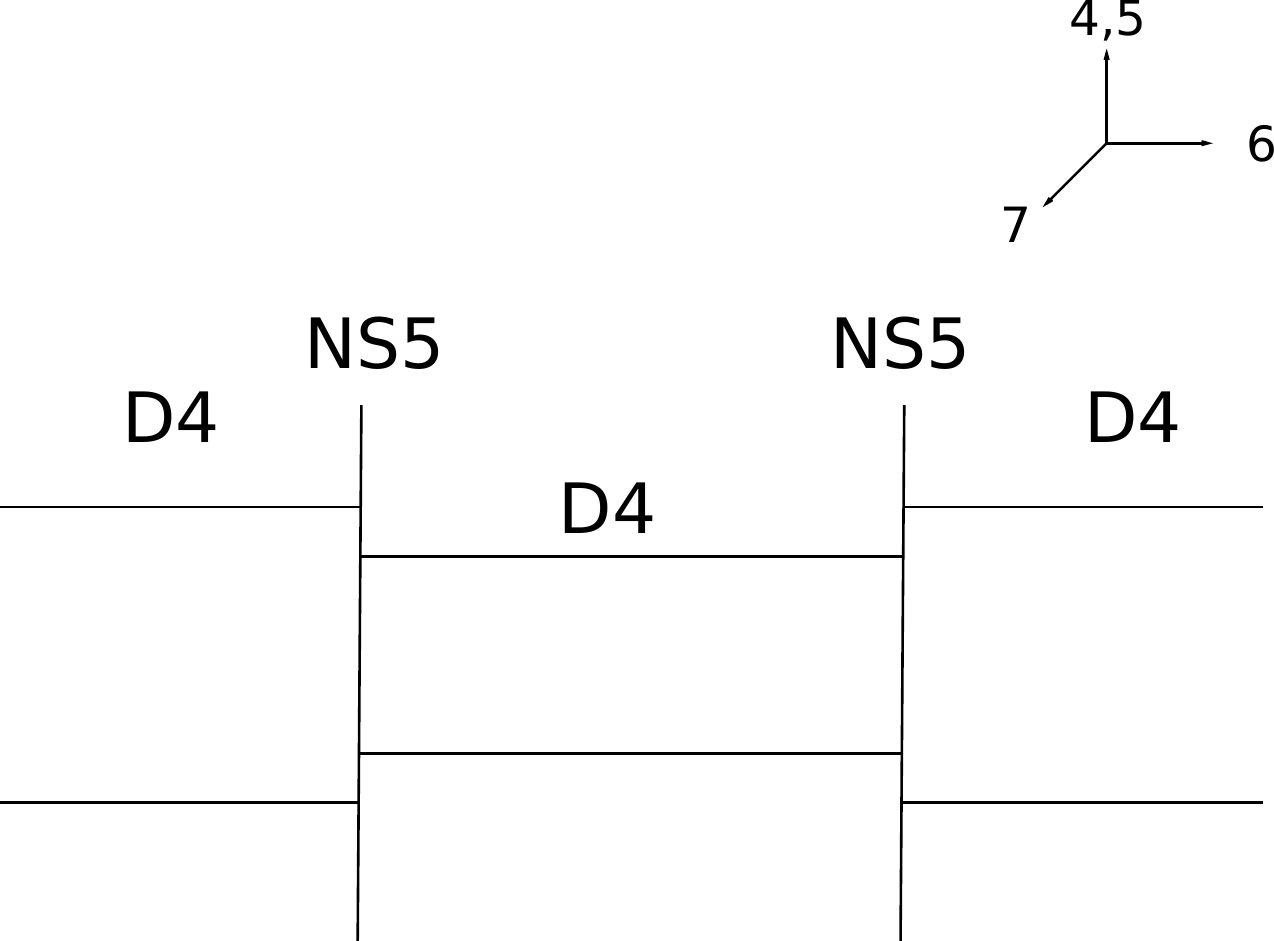}
\caption{A brane construction for  \texorpdfstring{$SU(2)$}{SU(2)} with \texorpdfstring{$N_f=4$}{Nf=4}, where all the four semi-infinite D4-branes end on two NS5-branes in pairs. This brane construction corresponds to the symmetric set up discussed in \autoref{sec:symbraneSU2single}.}
\label{fig:sym1copy}
\end{figure}

We begin in \autoref{sec:braneconstruction} by reviewing the brane construction of 4d $\N=2$ theories and their surface defects. In \autoref{sec:generalformulaGLSM} we review a different approach to
surface defects in Lagrangian 4d $\N=2$ theories, given in \cite{Gaiotto:2013sma}.
We then specialize to the $SU(2)$ theory with $N_f = 4$.
In \autoref{sec:asymbraneSU2single}, we discuss a surface defect which in the brane picture corresponds to the setup where all the semi-infinite D4-branes are put on one side of the NS5-branes, as shown in \autoref{fig:asym1copy}.
In \autoref{sec:symbraneSU2single}, we consider a surface defect which in the brane picture corresponds to the setup where the semi-infinite D4-branes are positioned symmetrically on both sides of the NS5-branes, as shown in \autoref{fig:sym1copy}.

\subsection{Brane constructions of \texorpdfstring{$\mathcal{N}=2$}{N=2} theories}\label{sec:braneconstruction}

4d $\mathcal{N}=2$ super Yang-Mills theory can be formulated by a brane construction in Type IIA string theory, as the worldvolume theory of a stack of D4-branes suspended between two NS5-branes. The gauge group is $SU(N)$, where $N$ is the number of parallel D4-branes. Fundamental hypermultiplets can be added using semi-infinite D4-branes ending on one of the NS5-branes. 

The NS5-branes are located classically at $x^7=x^8=x^9=0$ and a fixed value of $x^6$; the worldvolume of the D4-branes lives in the directions $x^0,x^1,x^2,x^3$ and $x^6$. In particular, each D4-branes ends on an NS5-brane in the $x^6$ direction on at least one side.

To describe the IR physics of the 4d theory, one can lift this brane configuration to M-theory, by adding an extra $S^1$ dimension with coordinate $x^{10}$. On the Coulomb branch of the theory, the whole brane configuration, consisting of D4-branes and NS5-branes in Type IIA, becomes a single M5-brane in M-theory. The surface spanned by this M5-brane embedded in the $x^{4,5,6,10}$ directions is related to the Seiberg-Witten curve of the class $S$ theory. More details can be found in \cite{Witten:1997sc,Gaiotto:2009hg}.

A 2d surface defect can be inserted in the 4d bulk theory by adding a D2-brane in the Type IIA theory, with one end on one of the NS5-branes, and the other end on a new NS5' brane whose worldvolume is
extended in the directions $x^{0,1,4,5,8,9}$. The worldvolume of the D2-brane is extended in the $x^{0,1,7}$ directions. The effective theory on the surface defect is an $\mathcal{N}=(2,2)$ supersymmetric $U(1)$ gauge theory. The complex scalar field of the vector multiplet on the defect is a combination of the directions $x^{4,5}$ in which the D2-brane can move. Chiral multiplets on the defect come from D2-D4 strings. In M-theory, the D2-brane lifts to an M2-brane, and the Fayet-Ilipoulos parameter is interpreted as the distance between the M2-brane and M5-brane in the complex $ix^6+x^{10}$ direction \cite{Hanany_1998,Alday:2009fs}.

Note that 
there are various D-brane systems which are different in the UV but 
flow to the same 4d $\N=2$ theory in the IR.
The above construction of surface defects can be applied to any of these D-brane systems;
these different constructions lead to genuinely different 
surface defects in the IR theory, as we will see below in examples.

\subsection{Coupling 2d \texorpdfstring{$\mathcal{N}=(2,2)$}{N=(2,2)} \texorpdfstring{$\mathbb{CP}^{N-1}$}{CP(N-1)} GLSMs to 4d \texorpdfstring{$\mathcal{N}=2$}{N=2} \texorpdfstring{$SU(N)$}{SU(N)} gauge theories}\label{sec:generalformulaGLSM}

4d $\mathcal{N}=2$ theories admit surface defects described by 2d $\mathcal{N}=(2,2)$ theories \cite{Gaiotto:2009fs}. In particular, in order to couple to a 4d $\mathcal{N}=2$ gauge theory with gauge group $G$, we can use a 2d $\mathcal{N}=(2,2)$ gauged linear sigma model whose flavor symmetry has a subgroup $G$, and introduce twisted chiral multiplets to gauge $G$. In this work, we will be mainly interested in $\mathbb{CP}^{N-1}$ sigma model defects coupled to bulk $SU(N)$ gauge theory with matter. In the gauged linear sigma model description, the $\mathbb{CP}^{N-1}$ sigma model has $N$ chiral multiplets transforming in the fundamental representation of $SU(N)$. The overall $U(1)$ symmetry is gauged by introducing a twisted chiral multiplet, whose scalar component is denoted as $\sigma$. One also turns on the FI parameter $t$. The bulk vector multiplet scalar vacuum expectation value $\Phi$ acts as a twisted mass for the $N$ chiral multiplets on the defect. After integrating out the chiral multiplets, one is left with an IR effective twisted superpotential $\WW$ for the twisted chiral scalar $\sigma$. 

The IR physics of the $2d$-$4d$ coupled system is determined by extremization of the effective twisted superpotential. The extremum equation is the twisted chiral ring relation for the multiple massive vacua of the theory. One can then reinterpret the twisted chiral ring relation as the Seiberg-Witten curve of the bulk theory, where the Seiberg-Witten differential $\lambda$ is identified as $\lambda = \sigma dt = \sigma \frac{dz}{z}$. Here $z=e^t$ is a complex parameter for the surface defect. The Seiberg-Witten curve is a fibration over the surface defect parameter space, whose sheets correspond to the different massive vacua of the surface defect \cite{Gaiotto:2009fs}.

As explained in \cite{Gaiotto:2013sma}, we could obtain the effective twisted superpotential in the following way. We start by weakly coupling the $SU(N)$ flavor symmetry of the $\mathbb{CP}^{N-1}$ sigma model to the bulk $SU(N)$ gauge fields. Semiclassically, eigenvalues of the 4d vector multiplet scalar $\Phi$, or equivalently the electric periods $a$, act as twisted mass parameters for the $N$ chiral multiplets living on the surface defect. Integrating out these chiral multiplets gives the following effective twisted superpotential:
\begin{equation}
2 \pi i\WW = t \sigma -\Tr\left[(\sigma + \Phi) \text{log}\left(\frac{\sigma +\Phi}{e}\right)\right]
\end{equation}

The four-dimensional gauge dynamics alters the twisted chiral ring of the surface defect. The quantum corrections could \ti{a priori} be hard to compute. In this case, we are lucky: they are captured by a well--studied object, the resolvent $T(\sigma)$. The resolvent is related to the low energy effective twisted superpotential as follows:
\begin{align}\label{seconddir}
T(\sigma):=\text{Tr}\frac{1}{\sigma+\Phi}=-2\pi \text{i}
\partial_\sigma^2\WW(\sigma).
\end{align}
$T(\sigma)$ can be computed using techniques in \cite{Cachazo:2002ry,Seiberg_2003,Cachazo_2003,Dijkgraaf_2002_1,Dijkgraaf_2002_2,nekrasov2003seibergwitten}.
In particular, the $T(\sigma)$ for the kind of defects we are interested in have been studied extensively in \cite{Gaiotto:2013sma}.
This gives us a way of extracting the surface defect chiral ring 
which is complementary to obtaining the 4d gauge theory from brane construction as reviewed in \autoref{sec:braneconstruction}. 

For example, suppose we couple the 
bulk $SU(N)$ theory with $N_f$ hypermultiplets to a surface defect described by the $\mathbb{CP}^{N-1}$ gauged linear sigma model. In this case $T(\sigma)$ has been worked out in \cite{Cachazo:2002ry,Gaiotto:2013sma}:
\begin{equation}\label{eqn:resolvent1}
T(\sigma)= \frac{B'(\sigma)}{2B(\sigma)} + \frac{2 B(\sigma) P_N'(\sigma) - B'(\sigma) P_N(\sigma)}{2 B(\sigma)\sqrt{P_N(\sigma)^2 - 4 \Lambda^{2N-N_f} B(\sigma)}} \, .
\end{equation}
Here $P_N(\sigma)$ is the characteristic polynomial of the vector multiplet scalar $\Phi$ for the $SU(N)$ gauge group, and $B(\sigma)$ is the characteristic polynomial of the background vector multiplet scalars for the $U(N_f)$ flavor group. 

In the following, we will also consider coupling a generalization of the $\mathbb{CP}^{N-1}$ surface defect to the bulk $SU(N)$ gauge theory with matter, as described in \cite{Gaiotto:2013sma}. This is inspired from the brane construction of the $\mathbb{CP}^{N-1}$ surface defect. Concretely, we let $N_f^+$ semi-infinite D4-branes end on one NS5-brane, while the remaining $N_f^-=N_f-N_f^+$ semi-infinite D4-branes end on the other NS5-brane. This splits the $N_f$ fundamental hypermultiplets into two groups, and breaks the $U(N_f)$ flavor symmetry to $U(N_f^+)\times U(N_f^-)$. Correspondingly, the flavor characteristic polynomial $B(\sigma)$ also splits:
\begin{equation}
\label{mafirstappear}
B(\sigma) = B_+(\sigma)B_-(\sigma)=\prod_{i=1}^{N_f^+}(\sigma-m_{+,i})\prod_{j=N_f-N_f^-+1}^{N_f}(\sigma-m_{-,j})= \prod_{a=1}^{N_f}(\sigma-m_a).
\end{equation}
Now we add extra fields to the 2d $\mathbb{CP}^{N-1}$ GLSM model; namely we add 2d chiral multiplets in the anti-fundamental of $U(N_f^+)$. These fields are coupled to the original 2d chiral multiplets in the GLSM and the bulk hypermultiplets via a superpotential term.
The contributions from these extra fields to the surface defect chiral ring can be absorbed
in a shift of the resolvent, replacing $T(\sigma)$ by $\hat{T}(\sigma)$ befined below:
\begin{equation}\label{eqn:resolvent2}
\hat T(\sigma) = \frac{B'(\sigma)}{2B(\sigma)} + \frac{2 B(\sigma) P_N'(\sigma) - B'(\sigma) P_N(\sigma)}{2 B(\sigma)\sqrt{P_N(\sigma)^2 - 4 \Lambda^{2N-N_f} B(\sigma)}} -\frac{B_+'(\sigma)}{B_+(\sigma)}.
\end{equation}

There is one subtle point: we cannot \ti{exactly} take over the results from \cite{Gaiotto:2013sma} to
our case. We consider a conformal theory with $2N = N_f$, and the most naive replacement
would be to replace $\Lambda^{2N-N_f}$ by $1$. We claim that the correct thing instead is to
replace it by a dimensionless quantity $f$ which depends on the exactly marginal coupling $q = e^{2\pi i \tau_{UV}}$. While we do not derive this rule from first principles here, we will find that if
we adopt this rule, the Seiberg-Witten curve we obtain
matches with the standard class $S$ Seiberg-Witten curve for $SU(2)$ with $N_f=4$ \cite{Gaiotto:2009hg,Tachikawa:2013kta}, when we make a particular choice for $f$; we discuss this point more in \autoref{sec:symbraneSU2single} below.

\subsection{Asymmetric construction for \texorpdfstring{$SU(2)$}{SU(2)} \texorpdfstring{$N_f=4$}{Nf=4}}\label{sec:asymbraneSU2single}
In this section we describe the pure $\mathbb{CP}^1$ surface defect coupled to the bulk $SU(2)$ $N_f=4$ theory. In the brane picture this corresponds to the asymmetric construction, where all four semi-infinite D4-branes end on one NS5-brane.

Thus we specialize the discussion above to the case of $N=2$, $N_f=4$. Then $P_N(\sigma)=\sigma^2-u$, where $u$ is the Coulomb branch parameter. We also simplify by setting 
all 4d hypermultiplet masses $m_a$ to zero. Thus
we take $B(\sigma)=\sigma^4$ and $B_+(\sigma)=1$. 

Integrating \eqref{eqn:resolvent1} to minimize the twisted effective superpotential gives the twisted chiral ring relation
\begin{equation}
0=t - \log{\left(\frac{P_N(\sigma) + \sqrt{P_N(\sigma)^2 - 4 f B(\sigma)}}{2}\right)}.
\end{equation}
The twisted chiral ring relation can be transformed into the class $S$ 
description of the Seiberg-Witten curve using the substitutions
\begin{align}
z &= e^t,\\
\lambda &= \sigma \, dt = \frac{\sigma}{z} \, dz,
\end{align}
which gives the curve as
\begin{align}
0 = \frac{u}{f z^3}+\frac{1}{f z^2}-\frac{\lambda ^2}{f z}+\lambda^4.
\end{align}

There are various class $S$ realizations of the $SU(2)$ theory with $N_f=4$. 
The particular one we found here appears to correspond to the theory of type 
$A_3$, reduced on a sphere with one regular and one irregular puncture. 
As far as we know, this particular class $S$ description of the theory 
has not been studied in detail; it would be interesting to do so.

\subsection{Symmetric construction for \texorpdfstring{$SU(2)$}{SU(2)} \texorpdfstring{$N_f=4$}{Nf=4}}\label{sec:symbraneSU2single}

Now we add 2d chiral multiplets in the anti-fundamental of the $U(2)\subset SO(8)$ bulk flavor symmetry, and couple them to the original chiral multiplets in the $\mathbb{CP}^1$ model and the bulk hypermultiplets. This corresponds to the brane construction where two semi-infinite D4-branes end on each of the two NS5-branes. This construction is what we will use for the doubled theory below.

In this case, we should replace the resolvent by \eqref{eqn:resolvent2}; equivalently, the relevant term appearing in $\partial_\sigma\WW$ becomes
 \begin{equation}\label{modifiedresolvent}
\log\left(\frac{P_N(\sigma)+\sqrt{P_N(\sigma)^2-4fB(\sigma)}}{2B_+(\sigma)}\right),
\end{equation}
where $B(\sigma)=\sigma^4$ and $B_+(\sigma)=\sigma^2$.
The resulting chiral ring equation is
\begin{equation}
\lambda^2+\frac{u}{z(f-z+z^2)}=0.
\end{equation}
This result matches the standard Seiberg-Witten curve for the $SU(2)$ $N_f=4$ theory \cite{Tachikawa:2013kta}. The above curve has four poles at $\{0,c_+,c_-,\infty\}$ where $c_{\pm}(f)=\frac{1}{2}(1\pm\sqrt{1-4f})$ are solutions to $f-z+z^2=0$. In order to match with the standard convention, we rescale $z$ to $z/c_+(f)$. Then, the four poles are at $\{0,1,q,\infty\}$ and $q=\frac{c_-(f)}{c_+(f)}$ is the UV gauge coupling. Substituting in the expressions of $c_\pm(f)$ and solve for $f$, we obtain
\begin{equation}\label{f_q}
f=\frac{q}{(1+q)^2}.
\end{equation}
We will use this substitution below when we study the doubled $SU(2)$ $N_f=4$ theory.

\section{The doubled \texorpdfstring{$SU(2)$}{SU(2)} \texorpdfstring{$N_f=4$}{Nf=4} theory} \label{chap3}
\subsection{Twisted chiral ring equation}
Surface defects in the pure $SU(2) \times SU(2)$ theory are studied in \cite{Gaiotto:2013sma}. In this section, we will similarly study a surface defect in the doubled $SU(2)$
$N_f = 4$ theory.
We gauge the $SU(2)\times SU(2)$ subgroup of the flavor symmetry group $U(4)$ of the 
gauged linear sigma model for  $\mathbb{CP}^3$. There is still a diagonal $U(1)$ commuting with the gauged subgroup. It corresponds to the 2d twisted mass $m$, which we set to zero.

The effective twisted superpotential on the surface defect is nearly decoupled,
with one term for each gauge group factor \cite{Gaiotto:2013sma}:
\begin{equation}
2\pi i \mathcal{W}(\sigma,\Phi,\tilde{\Phi})=t\sigma-\text{Tr}\left[(\sigma+\Phi)\text{log}\left(\frac{\sigma+\Phi}{e}\right)\right]-\text{Tr}\left[(\sigma+\tilde{\Phi})\text{log}\left(\frac{\sigma+\tilde{\Phi}}{e}\right)\right]
 \end{equation}
where $\Phi$ and $\tilde{\Phi}$ are the adjoint scalar fields in the two $SU(2)$ theories respectively.
The twisted chiral ring is obtained by setting $\partial_\sigma \cW$ to zero:
\begin{equation}
\label{twistedchiralringproduct}
t-\text{Tr}\log(\sigma+\Phi)-\text{Tr}\log(\sigma+\tilde{\Phi})=0.
\end{equation}

In \autoref{sec:asymbraneSU2single} and \autoref{sec:symbraneSU2single} we described two different 
surface defects in the $SU(2)$ $N_f=4$ theory, corresponding to two different Seiberg-Witten curves. It turns out to obtain the desired curve for the doubled $SU(2)$ $N_f=4$ theory, we need to use the analog of the surface defect in \autoref{sec:symbraneSU2single}, where the $SO(8)$ flavor symmetry for each copy of $SU(2)$ $N_f=4$ is broken to $U(2)\times U(2)$. We do not have a complete understanding of why we need to use the symmetric brane construction; one thing we can say is that the asymmetric one would be unlikely to work, since the defect in \autoref{sec:asymbraneSU2single} does not break the bulk $SO(8)$ flavor symmetry, while we expect flavor symmetry breaking from the class $S$ construction in \autoref{fig:example-C}. 

Now we use (\ref{modifiedresolvent}) for $\text{Tr}\log\left(\sigma+\Phi\right)$ in each copy of the 
$SU(2)$ $N_f=4$ theory. 
Then, the twisted chiral ring equation (\ref{twistedchiralringproduct}) becomes:
\begin{equation}
t-\log{\frac{P_N(\sigma)+\sqrt{P_N(\sigma)^2-4fB(\sigma)}}{2B_+(\sigma)}}-\log{\frac{P_N'(\sigma)+\sqrt{P_N'(\sigma)^2-4f'B(\sigma)}}{2B_+(\sigma)}}=0.
\end{equation}
Here $P_N=\sigma^2-u$ and $P_N'(\sigma) = \sigma^2-u'$, where $u$ and $u'$ are the Coulomb branch coordinates for the two decoupled theories, and as in \autoref{sec:symbraneSU2single} we take $B(\sigma)=\sigma^4$ and $B_+(\sigma)=\sigma^2$. To make formulas as compact as possible, here we keep the dimensionless parameters $f$ and $f'$. These parameters are related to the exactly marginal couplings $q$ and $q'$ of the two copies in the same way as in (\ref{f_q}).
 
Substituting $z=e^t$, the Seiberg-Witten differential is $\lambda dz=\sigma dt=\frac{\sigma}{z}dz$. We obtain the Seiberg-Witten curve 
\begin{equation} \label{eq:sw-curve-phi24}
	\lambda^4+\phi^{(2)}\lambda^2+\phi^{(4)}=0
\end{equation} with the following meromorphic differentials:
\begin{equation}\label{su4quiver}
\begin{split}
\phi^{(2)}&=\frac{u \left(z^2-2f'z+ff'\right)+u' \left(z^2-2fz+ff'\right)}{z\left(z^4-z^3+z^2\left(f+f'-2ff'\right)-ff'z+f^2f'^2\right) } \, dz^2 \, ,  \\
\phi^{(4)}&=\frac{\left(u z-u'f\right) \left(uf'-u' z\right)}{z^3\left(z^4-z^3+z^2\left(f+f'-2ff'\right)-ff'z+f^2f'^2\right) } \, dz^4 \, . 
\end{split}
\end{equation}
So we have found that the surface defect moduli space $\hC$ is a 6-punctured
sphere, with punctures at 
\begin{equation}
\begin{split}
&z_1=0,\quad z_2=\frac{1}{(1+q)(1+q')},\quad z_3=\frac{q}{(1+q)(1+q')},\\ &z_4=\frac{q'}{(1+q)(1+q')},\quad z_5=\frac{qq'}{(1+q)(1+q')},\quad z_6=\infty.
\end{split}
\end{equation} 
The Seiberg-Witten curve we obtained is 
a connected 4-fold cover of $\hC$, rather than
the disconnected union of two 2-fold covers. This is reasonable: although the two $SU(2)$ $N_f=4$ theories are decoupled, the surface defect is coupled to both of them.

Now, in accordance with our general strategy, we could
ask whether $\hC$ is the UV curve in an $A_3$ class $S$ realization
of the doubled $SU(2)$ $N_f=4$ theory. However, this is not the case. Indeed,
based on the pole orders of $\phi^{(2)}$ and $\phi^{(4)}$, we see that there are two maximal punctures at $z_1$ and $z_6$, and four minimal punctures at $z_2,z_3,z_4,z_5$; the corresponding
class $S$ theory is a linear $SU(4)$ quiver theory with three $SU(4)$ gauge nodes, bifundamental hypermultiplets between gauge nodes and fundamental hypermultiplets at two ends of the quiver. In particular, it is not the doubled
$SU(2)$ $N_f=4$ theory.

Nevertheless, we are getting close to our goal. We notice that the differentials in (\ref{su4quiver}) transform in a covariant way under a $\Z_2$ symmetry of the six-punctured sphere, 
given by
\begin{equation}
z \rightarrow \frac{qq'}{(1+q)^2(1+q')^2 z},
\end{equation}
as mentioned in \autoref{intro:an example}.
In \autoref{sec:productSU2} below, we will take the quotient by this $\Z_2$ symmetry and recover
the Seiberg-Witten curves corresponding to the class $S$ realization of the factorized theory
which was given in \cite{Chacaltana:2012ch}.
We discuss the physics of this quotient 
operation a bit more in \autoref{sec:quotient}.

\subsection{Seiberg-Witten curve for the doubled \texorpdfstring{$SU(2)$}{SU(2)} \texorpdfstring{$N_f=4$}{Nf=4} theory}\label{sec:productSU2}
To take the $\mathbb{Z}_2$ quotient of the surface defect parameter space $\hC$, we rewrite the differentials in (\ref{su4quiver}) in terms of the following $\Z_2$-invariant coordinate:
\begin{equation}
w=z+\frac{qq'}{(1+q)^2(1+q')^2z} \, .
\end{equation} 
The  meromorphic differentials become:
\begin{align}
\phi^{(2)}&=\frac{u \left(w-2 f'\right)+u' \left(w-2f\right)}{\left(w^2-w +f+f'-4ff'\right) \left(w^2 - 4ff'\right)} \, dw^2 \, , \label{quaddiffw}\\
\phi^{(4)}&=\frac{u'^2 f+u^2  f'- uu'w}{\left(w^2-4ff'\right)^2\left( w^2-w+f+f'-4ff'\right)} \, dw^4 \, . \label{quartdiffw}
\end{align}
Here again the dimensionless parameters $f,f'$ are related to the exactly marginal couplings $q,q'$ according to (\ref{f_q}). There are 5 poles located at 
\begin{equation*}
w_{1,2}=\mp\frac{2\sqrt{qq'}}{(1+q)(1+q')}, \quad 
w_3=\frac{q+q'}{(1+q)(1+q')},\quad
w_4=\frac{1+qq'}{(1+q)(1+q')},\quad 
w_5=\infty \, .
\end{equation*}
Analyzing pole orders of $\phi^{(2)}$ and $\phi^{(4)}$, we see there are two minimal untwisted punctures at $w_{3,4}$, two minimal twisted punctures (as defined in \cite{Chacaltana:2012ch,Chacaltana:2012zy}) at $w_{1,2}$, and one maximal untwisted puncture at $w=\infty$. This result matches with the twisted $A_3$ class $S$ construction from \cite{Chacaltana:2012ch}, depicted in \autoref{fig:example-C}. We remark that the two fixed points under the $\Z_2$ action become the locations of the minimal twisted punctures
on the quotient; this is not a surprise, as we will explain in \autoref{sec:quotient} below. 

We can also recover the two gauge couplings of the product theory from our construction; they can be expressed in terms of two cross-ratios for locations of the five punctures. Concretely we use the following two cross-ratios:
\begin{equation}\label{eqn:gaugecouplings}
\begin{split}
s_1&=\frac{(w_1-w_3)(w_2-w_5)}{(w_1-w_5)(w_2-w_3)}=\left(\frac{\sqrt{q}+\sqrt{q'}}{\sqrt{q}-\sqrt{q'}}\right)^2,\\
s_2&=\frac{(w_1-w_4)(w_2-w_5)}{(w_1-w_5)(w_2-w_4)}=\left(\frac{\sqrt{qq'}+1}{\sqrt{qq'}-1}\right)^2.
\end{split}
\end{equation}
The fact that $s_1$ and $s_2$ are complete squares signals that the gauge coupling moduli space is a four-fold covering of the complex structure moduli space for the five-punctured sphere, as described in \cite{Chacaltana:2012zy}.
To make the comparison more explicit, we take
\begin{equation}
x:=\frac{\sqrt{q}+\sqrt{q'}}{\sqrt{q}-\sqrt{q'}},\quad
y:=\frac{\sqrt{qq'}+1}{\sqrt{qq'}-1}
\end{equation}
where $(x,y)$ are coordinates on the four-fold covering. Then, the UV gauge couplings are given by
\begin{equation}
q=\frac{x-1}{x+1}\frac{y+1}{y-1},\quad
q'=\frac{x-1}{x+1}\frac{y-1}{y+1}.
\end{equation}
This is also exactly the same dependence as given in \cite{Chacaltana:2012zy}. Moreover, as explained in \cite{Chacaltana:2012zy}, there are certain deck transformations of the fourfold covering induced by $q\to 1/q$ and $q'\to 1/q'$. Concretely these deck transformations are generated by
\begin{equation}
\begin{split}
&(x,y)\to(-x,y)
\Longleftrightarrow (q,q')\to (1/q',1/q),\\
&(x,y)\to(x,-y)
\Longleftrightarrow
q\leftrightarrow q',\\
& (x,y)\to (y,x) \Longleftrightarrow (q,q')\to (1/q, q').
\end{split}
\end{equation}
In our construction, the $q\to 1/q$ symmetry is manifest: it arose from the substitution (\ref{f_q}). 

\subsection{Ultraviolet curve and parameter space for surface defect}\label{sec:quotient}

Here we motivate a little bit why the original surface defect parameter space is a double cover of the UV curve in the twisted $A_3$ realization of the two copies of $SU(2)$ with four flavors theory. 

First of all, we recall that 6d $(2,0)$ theory of type $\mathfrak{g}$ admits codimension-two defects. Due to lack of a path integral description of the 6d theory, an effective way to study these defects is by describing the singularities that they induce in the protected operators of the theory. Concretely, one could consider putting the 6d theory on $\R^{2,1}\times C \times S^1$ with codimension-two defects located at punctures on $C$. After compactifying on $S^1$, the codimension-two defects could be described by coupling a 3d $N=4$ superconformal field theory to the 5d $\N=2$ super Yang-Mills \cite{Gaiotto:2009hg,Gaiotto:2008sa,Gaiotto:2008ak,Chacaltana:2012zy}. To make connections with the usual setting to study class S theories, one further compactifies on $C$ with a partial topological twist, the moduli space of the resulting 3d theory is identified with the moduli space of solutions to Hitchin's equations on $C$, where the codimension-two defects specify the singular behavior of the solutions near their location on $C$ \cite{Gaiotto:2009hg}. In particular, locally the Higgs field $\Phi(z)$ takes the following form:
\begin{equation}
\Phi(z)\sim\Bigg(\frac{\Phi_{-1}}{z}+\cdots\Bigg) \, dz \, ,
\end{equation}
where $z$ is a local complex coordinate such that the defect sits at $z=0$. $\Phi_{-1}$ is an element in a nilpotent orbit of $\mathfrak{g}$. 

\begin{figure}
	\centering
	\includegraphics[width=300pt]{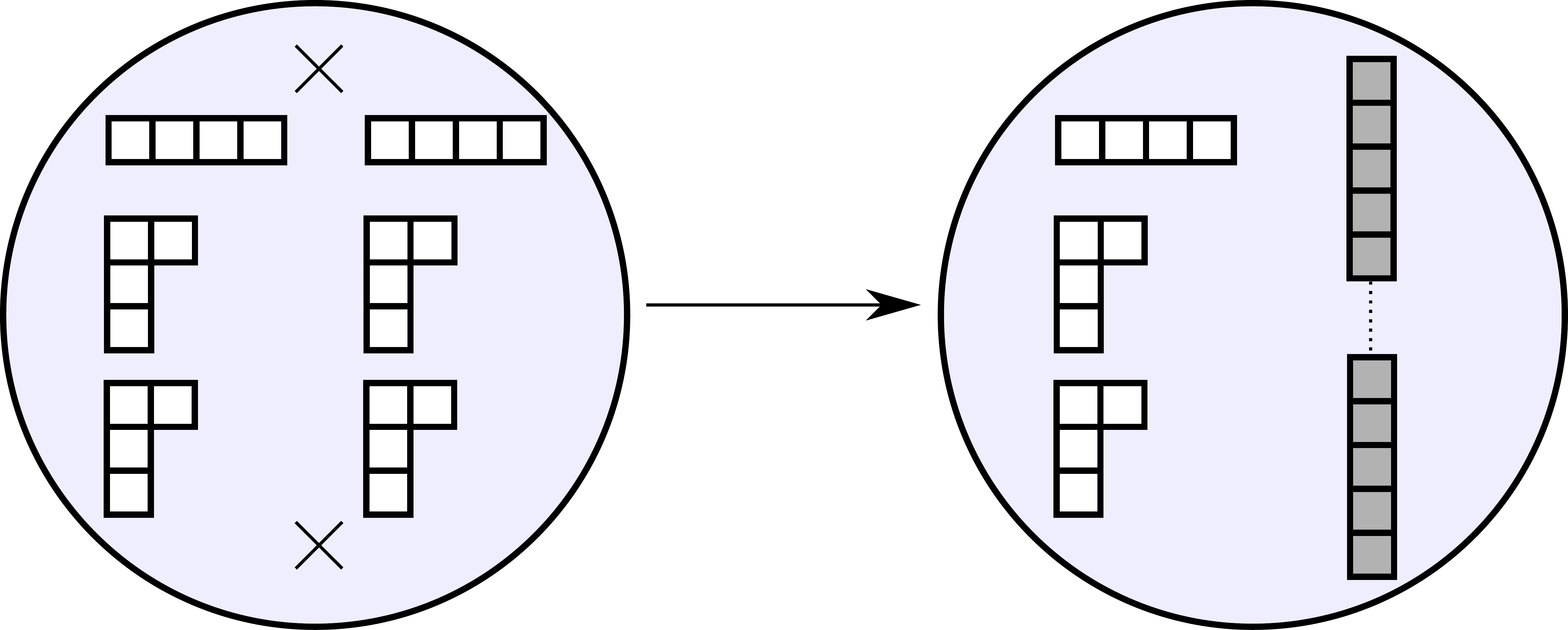}
	\caption{Illustration of the quotient action. Before the quotient, we have a six-punctured sphere with four minimal untwisted punctures (shown as the Young diagram $[3,1]$) and two maximal untwisted punctures (shown as the Young diagram $[1^4]$). Additionally we show the two fixed points of the quotient action as black crosses. After the quotient, we obtain a five-punctured sphere, where the two minimal twisted punctures are shown as the gray Young diagram $[5]$. There is a $\mathbb{Z}_2$ twist line (shown as a black dotted line) connecting these two twisted punctures.}
	\label{quotient}
\end{figure}

When $\mathfrak{g}$ has a $\Z_2$ outer automorphism $\mathfrak{o}$, there exists a twisted sector of codimension-two defects \cite{Chacaltana:2012zy}. Under the action of $\mathfrak{o}$, $\mathfrak{g}$ splits as a direct sum of the $\pm 1$ eigenspaces of $\mathfrak{o}$:
\begin{equation}
\mathfrak{g}=\mathfrak{g}_{1}+\mathfrak{g}_{-1}.
\end{equation}
In our case we have $\mathfrak{g}=A_{2N-1}$, $\mathfrak{g}_1=C_N$ and $\mathfrak{g}_{-1}=B_N$. In the neighborhood of a $\mathbb{Z}_2$-twisted puncture with a local complex coordinate $z$ where the defect is located at $z=0$, the Higgs field $\Phi(z)$ has the following expansion:
\begin{equation}\label{twistphi}
\Phi(z) \sim \Bigg(\frac{\Phi_{-1/2}}{z^{1/2}}+\frac{\Phi_{-1}}{z}+\cdots\Bigg) \, dz \, .
\end{equation}
Here $\Phi_{-1}$ is an element in a nilpotent orbit in $\mathfrak{g}_1$, and $\Phi_{-1/2}$ is a generic element in $\mathfrak{g}_{-1}$. Globally speaking, $\mathbb{Z}_2$-twisted punctures always appear in pairs, with $\mathbb{Z}_2$ twist lines connecting the two punctures within each pair, see \autoref{quotient} for an example. 

The square root in (\ref{twistphi}) suggests that the Higgs field is well-defined on a double cover of the UV curve $C$. It is then natural to expect the surface defect parameter space to be a double cover of $C$. This is indeed what happened in our example, as described in \autoref{sec:productSU2} and \autoref{sec:quotient}. After taking the $\Z_2$ quotient, we obtain the Seiberg-Witten curve for two copies of $SU(2)$ with $N_f=4$, realized in the $\Z_2$-twisted $A_3$ theory 
\cite{Chacaltana:2012ch}. In particular, the two fixed points under the $\Z_2$ action become the locations of two minimal twisted punctures. There is a $\Z_2$ twist line connecting these two punctures, which could also be understood as a choice of branch cut after the $\Z_2$ quotient action. This procedure is illustrated in \autoref{quotient}.

\section{Factorization of the Coulomb branch}\label{chapter:CBfactorization}

As described in \autoref{sec:CBfactorize}, in a factorized theory on its Coulomb branch, the periods $Z_\gamma(u)$ of the Seiberg-Witten differential must decompose in the form \eqref{eqn:Zgammadecom}. In this section we check this decomposition in two examples of factorized class $S$ theories: in \autoref{CBfactorizationSU2} we study the Coulomb branch of the doubled $SU(2)$ $N_f=4$ theory, while in \autoref{CBfactorizeE6} we turn to the doubled rank-$1$ $E_6$ Minahan-Nemeschansky theory. For simplicity, we will focus on special regions of the Coulomb branch.

\subsection{Coulomb branch study of the doubled \texorpdfstring{$SU(2)$}{SU(2)} \texorpdfstring{$N_f=4$}{Nf=4} theory}\label{CBfactorizationSU2}

\begin{figure}[tbp]
	\centering 
	\includegraphics[width=0.8\textwidth]{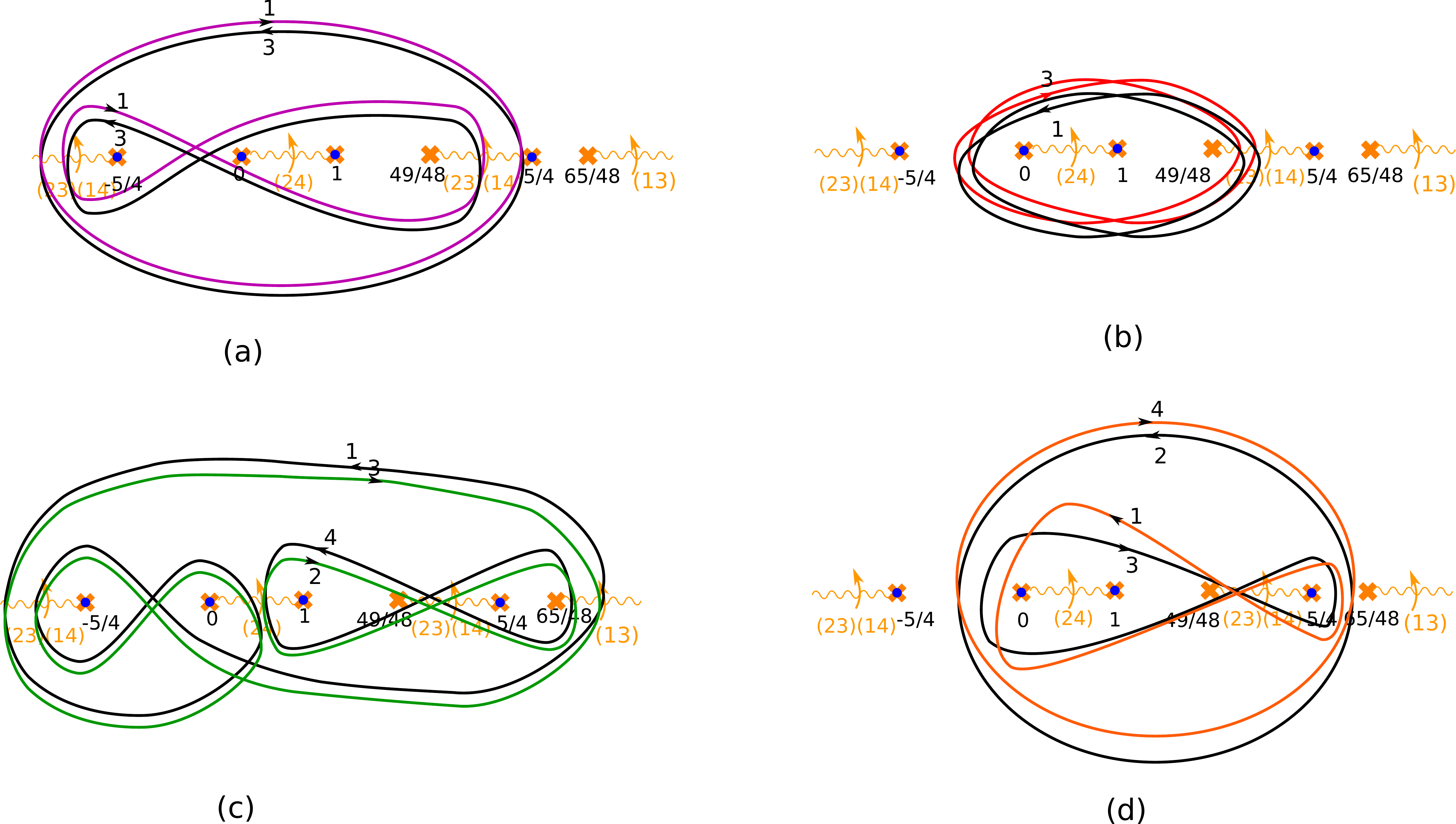}
	\caption{\label{fig:decomposition}{Basis vectors for the EM charge lattice in the doubled $SU(2)$ $N_f=4$ theory, realized as homology classes on the Seiberg-Witten curve $\overline\Sigma$. These charges have $-\langle\gamma_a,\gamma_d\rangle=\langle\gamma_d,\gamma_a\rangle=-\langle\gamma_c,\gamma_b\rangle=\langle\gamma_b,\gamma_c\rangle=8$. We set $f=5/16,\ f'=5/4,\ u=1,\ u'=3$. The poles are at $z=-5/4,0,1,5/4$, and there are two other branch points at $z=49/48$ and $z=65/48$. We construct the cycles with $\lambda\rightarrow -\lambda$ symmetry; thus for each strand on sheet $i$, there is a strand on sheet $i+2$ mod $4$ going in the opposite direction.}}
\end{figure}

First we consider the doubled $SU(2)$ $N_f=4$ theory, for which the Seiberg-Witten curve $\Sigma$
is given by \eqref{eq:sw-curve-phi24} with the differentials \eqref{quaddiffw}, \eqref{quartdiffw}. After filling in the punctures we obatin
$\overline\Sigma$ which is a genus 3 surface. The full $H_1(\overline{\Sigma},\mathbb{Z})$ has rank $6$, but the fact that $\phi^{(3)}=0$ results in a symmetry $\lambda\rightarrow-\lambda$; restricting
to anti-invariant cycles under this involution gives a rank 4 sublattice $\Gamma \subset H_1(\overline{\Sigma},\mathbb{Z})$. 4 linearly independent classes in $\Gamma$ are indicated in \autoref{fig:decomposition}; although we have not carefully developed the theory of charge lattices in
twisted $A_n$ class $S$ theories, we suspect that these cycles (up to scalar multiple) form a basis for the
charge lattice in this theory.

The central charge corresponding to an EM charge $\gamma$ is the integral $Z_\gamma=\frac{1}{\pi}\oint_\gamma \lambda$. We choose the gauge couplings to be $(f,f')=(5/16,5/4)$ and vary the Coulomb branch parameters in the neighborhood of $(u,u')=(1,3)$. We choose the four linearly independent charges shown in \autoref{fig:decomposition} and calculate the central charges numerically; we find

\begin{equation}
Z_a \approx - 5.09 i \sqrt{u'},
\end{equation}
\begin{equation}
Z_b \approx 5.42 \sqrt{u},
\end{equation}
\begin{equation}
Z_c \approx 9.81 i \sqrt{u}, 
\end{equation}
\begin{equation}
Z_d \approx -4.09  \sqrt{u'}.
\end{equation}
Moreover, varying $(f,f')$ slightly we see that $Z_{b,c}$ depend only on $f$ and
$Z_{a,d}$ only on $f'$.
The most general central charges are linear combinations of the four above, and thus have the 
form $g(f)\sqrt{u}+g'(f')\sqrt{u'}$. This reflects the expected factorization into two
copies of the $SU(2)$ $N_f=4$ theory (recall that the 
central charges in that theory have the form $g(f)\sqrt{u}$, where $g(f(q))$ is a function of the coupling and $u$ is the Coulomb branch parameter). 

\subsection{Coulomb branch study of the doubled \texorpdfstring{$E_6$}{E6} Minahan-Nemeschansky theory}\label{CBfactorizeE6}

\begin{figure}
	\centering
	\includegraphics[width=108pt]{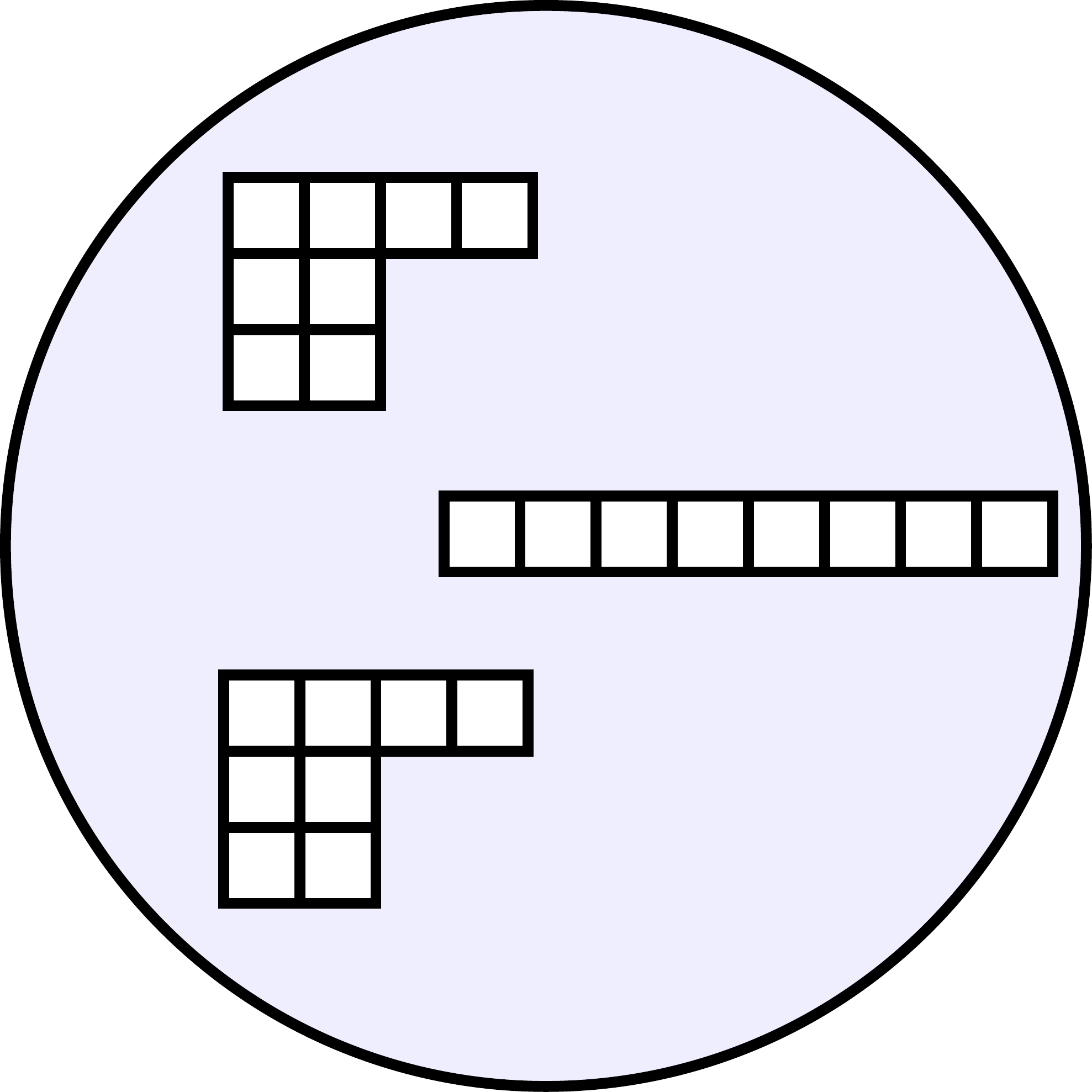}
	\caption{An interacting fixture in the $D_4$ theory realizing the doubled rank-1 $E_6$ Minahan-Nemeschansky theory.}
	\label{fig:D4eg}
\end{figure}

In this section we turn our attention to another interesting factorized SCFT, realized by compactifying the 6d $(2,0)$ theory of type $D_4$ on the three-punctured sphere shown in \autoref{fig:D4eg}. The corresponding 4d $\mathcal{N}=2$ SCFT was identified in \cite{Chacaltana:2011ze} as the direct sum of two copies of the rank-$1$ Minahan-Nemeschansky $E_6$ SCFT. In the following we will simply denote this theory as $\mathcal{T}_{D_4}$. Since the rank-1 $E_6$ theory is non-Lagrangian, we cannot use the machinery in \autoref{chap3} directly to understand this factorization. However, we can check
directly that the Coulomb branch indeed factorizes. 

We first review the Coulomb branch of the rank-1 Minahan-Nemeschansky $E_6$ SCFT. A realization of this theory in class $S$ is given by a fixture in the $A_2$ theory with three full punctures, often called the $T_3$ theory. $T_3$ has a one-dimensional Coulomb branch, parameterized by $u\in\mathbb{C}$. The period integrals are proportional to $u^{1/3}$, consistent with the fact that $u$ has scaling dimension 3. Below we will analyze the Seiberg-Witten geometry of $\mathcal{T}_{D_4}$ and discover that the period integrals and charge lattice of $\mathcal{T}_{D_4}$ indeed have the desired factorized behavior.\footnote{A preliminary version of this analysis appeared in \cite{Yan2018}.}

\begin{figure}
	\centering
	\includegraphics[scale=0.3]{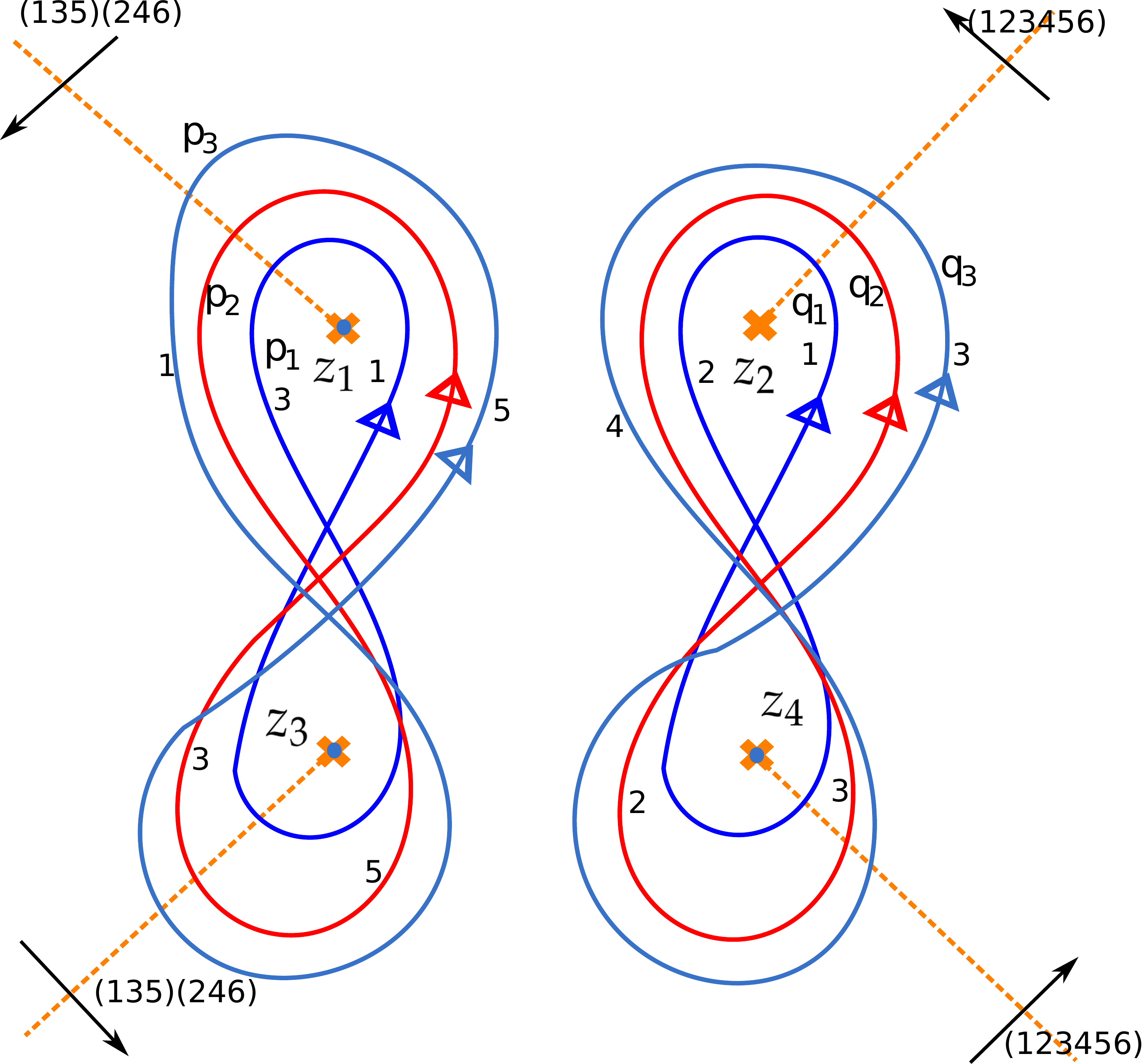}
	\caption{Choice of branch cuts for $\overline{\Sigma'}$ and cycles in homology classes $p_i$ and $q_i$ ($i=1,2,3$) in $H_1(\overline{\Sigma'},\mathbb{Z})$. We pick Coulomb branch parameters such that the four branch points are at $z_1=0$, $z_2=z_p$, $z_3=1$ and $z_4=\infty$, as shown in this figure. The four branch cuts are represented by wavy lines and they meet at some point on $\mathbb{CP}^1$. Each branch cut is labeled with the corresponding sheet permutation and the direction to do the gluing. }
	\label{D4cuts}
\end{figure}

The UV curve is $C=\mathbb{CP}^1\setminus \{0,1,\infty\}$, with the punctures at $z=0$ and $z=1$ labeled by the $D$-partition $[3,3,1,1]$, and the puncture at $z=\infty$ a full puncture.
The meromorphic $k$-differentials parameterizing the Coulomb branch 
in the $D_4$ theory are $\phi^{(2)},\phi^{(4)},\phi^{(6)}$ and the Pfaffian $\tilde{\phi}$. The pole structures and constraints at the three punctures 
(as worked out in \cite{Chacaltana:2011ze}) 
imply that the only nonvanishing differential is $\phi^{(6)}$, which has at most fourth-order poles at $z_{1,2}=0,1$, and at most a fifth-order pole at $z_3=\infty$. Moreover, the leading pole coefficients of $\phi^{(6)}$ at $z_{1,2}=0,1$ satisfy constraints of the form $c_{1,2}=(v_{1,2})^2$, where $v_{1,2}$ parametrizes the Coulomb branch \cite{Chacaltana:2011ze,Chacaltana:2012zy}.\footnote{This constraint is necessary for the purpose of obtaining the correct physical quantities in the theory, such as the graded Coulomb branch dimension etc. \cite{Chacaltana:2012zy}} As a result, the Coulomb branch of $\mathcal{T}_{D_4}$ is a four-fold covering of the space of meromorphic differentials with required pole constraints. The Seiberg-Witten curve $\Sigma\subset T^*C$ is then given by
\begin{equation}
\lambda^2(\lambda^6-\phi^{(6)})=0,
\end{equation} 
where
\begin{equation}\label{phi6}
\phi^{(6)}=\frac{v_1^2+(v_2^2-v_1^2)z}{z^4(z-1)^4}dz^6.
\end{equation}
$\lambda$ is the Seiberg-Witten differential, concretely in terms of coordinates $(x,z)$ on $T^*C$, $\lambda = x dz$. Two sheets of $\Sigma$ are trivial coverings of $C$; the other six sheets form a branched covering of $C$, with a branch point at $z_p:=\frac{v_1^2}{v_1^2-v_2^2}$. Denote this branched covering as $\Sigma'$ and fill in the punctures we obtain a genus 4 curve $\overline{\Sigma'}$; $\overline{\Sigma'}$ is a branched covering of $\mathbb{CP}^1$ with four branch points at $0,1,\infty,z_p$. Concretely, we can construct $\overline{\Sigma'}$ by gluing together six sheets along branch cuts as shown in \autoref{D4cuts}, where the six sheets are labeled by the six possible choices of sixth root of $\phi^{(6)}$. 

If the Coulomb branch of $\mathcal{T}_{D_4}$ has the desired factorization, then for any homology class $\gamma\in H_1(\overline{\Sigma'},\mathbb{Z})$, the corresponding period integral must take the form
\begin{equation}\label{period}
\oint_\gamma\lambda=A u_1^{1/3}+B u_2^{1/3},
\end{equation}
where $A,B\in\mathbb{C}$ and $u_{1,2}$ are some functions of $v_{1,2}$. We would like to confirm this property and identify $u_{1,2}$. For convenience, in the following we fix $\text{Re}(z_p)>1$. The analysis is valid in other regions by analytic continuation.

First let us look at the cycle $p_1$ depicted in \autoref{D4cuts}. We have
\begin{align*}
\oint_{p_1}\lambda  &=\big(\text{e}^{2\pi\text{i}/3}-1\big)\int_0^1\frac{(v_2^2-v_1^2)^{1/6}(z-z_p)^{1/6}}{z^{2/3}(z-1)^{2/3}}dz\\
&=9\big(1+\text{e}^{2\pi\text{i}/6}\big)\pi^{-\frac{1}{2}}\Gamma\Big(\frac{4}{3}\Big)\Gamma\Big(\frac{7}{6}\Big)\Big(\sqrt{v_1^2}+\sqrt{v_2^2}\Big)^{1/3}.
\end{align*}
Similarly for the cycle $q_1$ depicted in \autoref{D4cuts},
\begin{equation*}
\oint_{q_1}\lambda =\big(\text{e}^{2\pi\text{i}/6}-1\big)\int_{z_p}^\infty\frac{(v_2^2-v_1^2)^{1/6}(z-z_p)^{1/6}}{z^{2/3}(1-z)^{2/3}}dz \sim \Big(\sqrt{v_1^2}-\sqrt{v_2^2}\Big)^{1/3}.
\end{equation*}
Similarly one can argue that for any homology class $\gamma$ the period integral takes the form of (\ref{period}), with
\begin{equation}\label{CBparam}
u_1=\sqrt{v_1^2}+\sqrt{v_2^2}, \quad u_2=\sqrt{v_1^2}-\sqrt{v_2^2}.
\end{equation}
This reduces to showing that the integrals between branch points take the desired form, which is indeed the case. For example,
\begin{equation*}
\begin{split}
\int_1^\infty \frac{(v_2^2-v_1^2)^{1/6}(z-z_p)^{1/6}}{z^{2/3}(1-z)^{2/3}}dz&\sim (v_2^2-v_1^2)^{1/6}{}_2F_1\Big(-\frac{1}{6},\frac{1}{6};\frac{1}{2};z_p\Big)\\
&=(v_2^2-v_1^2)^{1/6}\text{cos}\Bigg(\frac{1}{3}\text{arcsin}\sqrt{\frac{v_1^2}{v_1^2-v_2^2}}\Bigg)\\
& \sim A\Big(\sqrt{v_1^2}+\sqrt{v_2^2}\Big)^{1/3}+B \Big(\sqrt{v_1^2}-\sqrt{v_2^2}\Big)^{1/3},
\end{split}
\end{equation*}
where $A,B\in\mathbb{C}$.\footnote{Here we have used $\text{cos}(3\alpha)=4\text{cos}^3\alpha-3\text{cos}\alpha$.}

The appearance of $\sqrt{v_i^2}$ above requires some care with branches. To exhibit the actual dependence of the periods on $v_{1,2}$, we consider numerical examples where $v_{1,2}$ are positive real numbers with $v_1>v_2$. Then we find 
\begin{align*}
\oint_{p_1}\lambda&=(-6.31+3.64\text{i})\big(v_1+v_2\big)^{1/3},\\
\oint_{q_1}\lambda&=(6.31+3.64\text{i})\big(v_1-v_2\big)^{1/3}.
\end{align*}
Thus we get the expected factorization, with
\begin{equation}
	u_1 = v_1 + v_2, \quad u_2 = v_1 - v_2.
\end{equation}

To go further we now take a closer look at the precise charge lattice of the theory.
Seiberg-Witten curves of type $D$ have a $\mathbb{Z}_2$ symmetry which takes $\lambda \to -\lambda$.
This symmetry acts on $H^1(\overline{\Sigma'},\Z)$  as the involution 
$\sigma$ which shifts the sheet number by $+3$ (mod 6); thus we have
\begin{equation}
\oint_p \lambda = -\oint_{\sigma(p)} \lambda, \quad p\in H^1(\overline{\Sigma'},\Z).
\end{equation}

In \autoref{D4cuts} we show three cycles in homology classes $p_1$, $p_2$, $p_3$ with
\begin{equation}
p_1+p_2+p_3=0, \quad 
\langle p_1,p_2 \rangle=\langle p_2, p_3\rangle=\langle p_3,p_1\rangle =1,
\end{equation}
where $\langle \cdot , \cdot \rangle$ denotes the intersection pairing in $H^1(\overline{\Sigma'},\Z)$. Now we consider the cycles $\{\gamma_1,\gamma_2,\gamma_3\}$ given by
\begin{equation}
\gamma_i=p_i-\sigma(p_i),
\end{equation}
which satisfy
\begin{equation}
\gamma_1+\gamma_2+\gamma_3=0, \quad 
\langle \gamma_1,\gamma_2 \rangle_{\text{DSZ}}=\langle \gamma_2, \gamma_3\rangle_{\text{DSZ}}=\langle \gamma_3,\gamma_1\rangle_{\text{DSZ}} =1.
\end{equation}
Here $\langle \cdot , \cdot \rangle_{\text{DSZ}}$ denotes the Dirac-Schwinger-Zwanziger pairing in the charge lattice of a class S theory of type $D$, which is given by the corresponding intersection pairing in $H^1(\overline{\Sigma'},\Z)$ divided by a factor of $2$ \cite{Longhi:2016rjt}. 
Moreover the central charges $Z_{\gamma_i}$ satisfy
\begin{equation}
Z_{\gamma_2}=w Z_{\gamma_1}, \quad Z_{\gamma_3}=w^2 Z_{\gamma_1},
\end{equation}
where $w=\text{e}^{2\pi\text{i}/3}$. This matches with the 
structure of the charge lattice in the rank-$1$ $E_6$ Minahan-Nemeschansky theory. 

Similarly we define $\{\gamma'_i\}$ using the classes $q_1$, $q_2$ and $q_3$ shown in \autoref{D4cuts}:
\begin{equation}
\gamma'_1=q_1-\sigma(q_1), \quad 
\gamma'_2=q_3-\sigma(q_3), \quad
\gamma'_3=\sigma(q_2)-q_2,
\end{equation}
which again satisfy
\begin{equation}
\langle \gamma'_1,\gamma'_2 \rangle_{\text{DSZ}}=\langle \gamma'_2, \gamma'_3\rangle_{\text{DSZ}}=\langle \gamma'_3,\gamma'_1\rangle_{\text{DSZ}} =1, \quad Z_{\gamma'_1}=w^{-1} Z_{\gamma'_2}=w^{-2}Z_{\gamma'_3}.
\end{equation}
Moreover the pairing between $\gamma_i$ and $\gamma_j'$ is always zero.

Thus we have found that the charge lattice of $\mathcal{T}_{D_4}$ factorizes into a product of two lattices, each isomorphic to the charge lattice in the rank-$1$ $E_6$ Minahan-Nemeschansky theory,
with the correct behavior of the central charges, as desired.

In \cite{Chacaltana:2011ze} the authors discovered another two interacting SCFTs closely related to the product $E_6$ theory. The Coulomb branches of those theories are quotients of the Coulomb branch of $\mathcal{T}_{D_4}$ discussed here. For example, the Coulomb branch for one of those theories (the $F_4\times SU(2)^2$ SCFT) is parametrized by $v_{1,2}^2$ instead of $v_{1,2}$. It would be interesting to adapt our analysis to obtain the charge lattice for this theory.\footnote{We thank Jacques Distler for pointing out this question.}

\bibliographystyle{utphys}
\bibliography{factorized-paper}

\end{document}